\documentclass[3p,times,procedia]{elsarticle}
\usepackage{nupha_ecrc}

\volume{00}
\firstpage{1}
\journalname{Nuclear Physics A}
\runauth{}
\jid{nupha}
\jnltitlelogo{Nuclear Physics A}
\usepackage{amssymb}
\usepackage[figuresright]{rotating}
\begin{document}
\begin{frontmatter}
\dochead{XXVIIth International Conference on Ultrarelativistic Nucleus-Nucleus Collisions\\ (Quark Matter 2018)}
\title{Measurements of directed and elliptic flow for $D^{0}$ and $\overline{D^{0}}$
mesons using the STAR detector at RHIC}

\author{ Subhash Singha (for the STAR Collaboration) \footnote{A list of members of the STAR Collaboration and acknowledgements can be found at the end of this issue.
S.S. acknowledges financial support from DOE project (Grant No. DE-FG02-89ER40531), USA.}
}
\address{Department of Physics, Kent State University, Ohio 44242, USA \\
subhash@rcf.rhic.bnl.gov
}

\begin{abstract}
We report on the first evidence for a non-zero rapidity-odd directed flow
($v_{1}$) for $D^{0}$ and $\overline{D^{0}}$ mesons in 10-80\% centrality Au+Au collisions at $\sqrt{s_{\rm NN}}$ =
200~GeV measured with the STAR detector at RHIC. The slope of the $v_1$ rapidity dependence 
($dv_{1}/dy$) averaged over $D^{0}$ and $\overline{D^{0}}$ mesons is  -0.081
$\pm$ 0.021 $\pm$ 0.017, while that of charged kaons is -0.0030 $\pm$
0.0001 $\pm$ 0.0002, suggesting significantly larger slope for the
$D^{0}$ mesons. Models indicate that the large $dv_{1}/dy$ of $D^{0}$ mesons is 
sensitive to the initially tilted source. We also present a new measurement of the $D^{0}$ meson
elliptic flow ($v_2$) as a function of transverse momentum ($p_{T}$) in Au+Au collisions at
$\sqrt{s_{\rm NN}}$ = 200~GeV with an improved precision with respect
to the previously published results.The $D^{0}$ $v_{2}$ results are
compared to those of light-flavor hadrons to test the number-of-constituent-quark (NCQ) scaling. Both the $v_{1}$ and
$v_{2}$ results are compared to recent hydrodynamic and transport model calculations.
\end{abstract}
\begin{keyword}
relativistic heavy-ion collisions \sep heavy flavor \sep directed flow \sep elliptic flow
\end{keyword}
\end{frontmatter}
\section{Introduction}
\label{intro}
Heavy quarks play a crucial role in probing the Quark Gluon Plasma (QGP) phase because
their masses are significantly larger than the typical temperature achieved in the medium. The production of
heavy quarks occurs mainly during the primordial stage of heavy-ion
collisions before the QGP is formed. As a consequence, they experience the entire evolution of the
system and can be used to access information concerning the early time
dynamics~\cite{rapp}. A recent hydrodynamic model~\cite{chatt_bozek_dv1_prl}, which incorporates Langevin
dynamics for heavy quarks combined with an initial tilt of the source~\cite{tilt_chv1_bozek},
predicts a relatively larger $v_{1}$ for heavy flavors compared to the
light ones. The model demonstrates the sensitivity of the $D$-meson
$v_{1}$ slope to the initially tilted geometry and the interaction
between charm quarks and the medium. Furthermore, another
model~\cite{das_greco_dv1_em} predicts that the transient
electromagnetic (EM) field generated in heavy-ion collisions can
induce opposite $v_{1}$ for charm ($c$) and anti-charm ($\bar{c}$)
quarks. Such an  EM-field-induced $v_{1}$ for hadrons containing heavy
quarks is predicted to be several orders of  magnitude larger than
that for light-flavor hadrons~\cite{gursoy_em}. Thus, the separate
measurement of $v_1$ for $D^{0}$ and $\overline{D^0}$ can offer
insight into the early-time EM fields.

Recent measurements at RHIC, based on 2014 data, have shown that $D^{0}$
mesons in minimum-bias and mid-central heavy-ion collisions exhibit
significant elliptic flow~\cite{star_d0_v2_ncq}. The flow magnitude follows 
the same number-of-constituent-quark (NCQ) scaling pattern as observed for light-flavor hadrons in
mid-central collisions. It is of particular interest to measure the centrality 
dependence of these observables and to test the NCQ scaling for charmed
hadrons in different centrality classes. During 2016, STAR~\cite{starnim} collected an additional
sample of Au+Au collisions at  $\sqrt{s_{\rm NN}}$ 200~GeV using 
the Heavy Flavor Tracker (HFT)~\cite{hft_tdr, hft_nim} detector. An
improved precision for the anisotropic flow measurements of heavy-flavor 
hadrons has been achieved by combining the data samples collected during 2014 and 2016 allowing more quantitative studies of the QGP properties.
\section{Analysis details}
\label{ana}
Minimum-bias events are defined by a coincidence of signals in the east and west Vertex 
Position Detectors (VPD)~\cite{VPD} located at pseudorapidity $4.4 < |\eta| < 4.9$.  
The collision centrality is determined from the number of charged particles within
$|\eta| < 0.5$ and corrected for triggering efficiency using a Monte Carlo Glauber 
simulation~\cite{centrality_glauber}. The $D^{0}$ and $\overline{D^{0}}$  mesons are 
reconstructed via their hadronic decay channels: $D^{0} \,(\overline{D^{0}}) \rightarrow
K^{-}\pi^{+} \,(K^{+}\pi^{-})$ (branching ratio: 3.89$\%$, c$\tau \sim$ 123 $\mu$m)~\cite{pdg}
by utilizing the Time Projection Chamber (TPC)~\cite{startpc} along with the HFT. 
Good-quality tracks with $p_{T} >$ 0.6 GeV/$c$ and $|\eta| < 1$ are
ensured by requiring a minimum of 20 TPC hits (out of possible 45),
and with at least one hit in each layer of the Intermediate Silicon
Tracker (IST) and PiXeL (PXL) components of the HFT~\cite{star_d0_v2_ncq}. The identification of
$D^{0}$ decay daughters is based on the specific ionization energy
loss ($dE/dx$) in the  TPC and on the velocity of
particles ($1/\beta$) measured by the Time of Flight (TOF)~\cite{TOF}
detector. To reduce the background and enhance the signal-to-background ratio, topological
variable cuts are optimized using the Toolkit for Multivariate Data
Analysis (TMVA) package~\cite{tmva, star_d0_v2_ncq}. The first-order
event plane azimuthal angle ($\Psi_{1}$) is reconstructed using the
Zero-Degree Calorimeter Shower Maximum Detectors (ZDC-SMDs)~\cite{zdc-thesis}. The ZDC-SMDs ($|\eta| > 6.3$) are
separated by about five units in pseudo rapidity from the TPC and the
HFT. This separation reduces significantly the possible non-flow effects in
$v_1$ measurements. The second-order event plane ($\Psi_{2}$) is
reconstructed from tracks measured in the TPC. To suppress the
non-flow effects in the $v_{2}$ measurements, only tracks that are in
the opposite rapidity hemisphere with at least $\Delta \eta > 0.05$
with respect to the reconstructed $D^{0}$, are employed for the
$\Psi_{2}$ reconstruction. The $v_{1}$ and $v_{2}$ coefficients are calculated using the
event-plane  method~\cite{methods} measuring the $D^{0}$ yields in
different azimuthal intervals defined with respect to the event plane
angle  ($\phi - \Psi_{n}$). The $D^{0}$ yields are weighted by the
inverse of the reconstruction efficiency $\times$ acceptance for each
interval of collision centrality. The observed $v_{n}$ is then
calculated by fitting the azimuthal dependence of the $D^{0}$ yield using
the  function $p_{0} (1 + 2\;v_n^{\rm obs} \cos [n\; (\phi - \Psi_n)])$. 
The resolution-corrected $v_n$ is then obtained by dividing $v_n^{\rm obs}$ by the 
event-plane resolution corresponding to $\Psi_n$~\cite{ep_resol}.
\section{Results}
\label{results}
\begin{figure}[!htb]
\begin{center}
\includegraphics[scale=0.35]{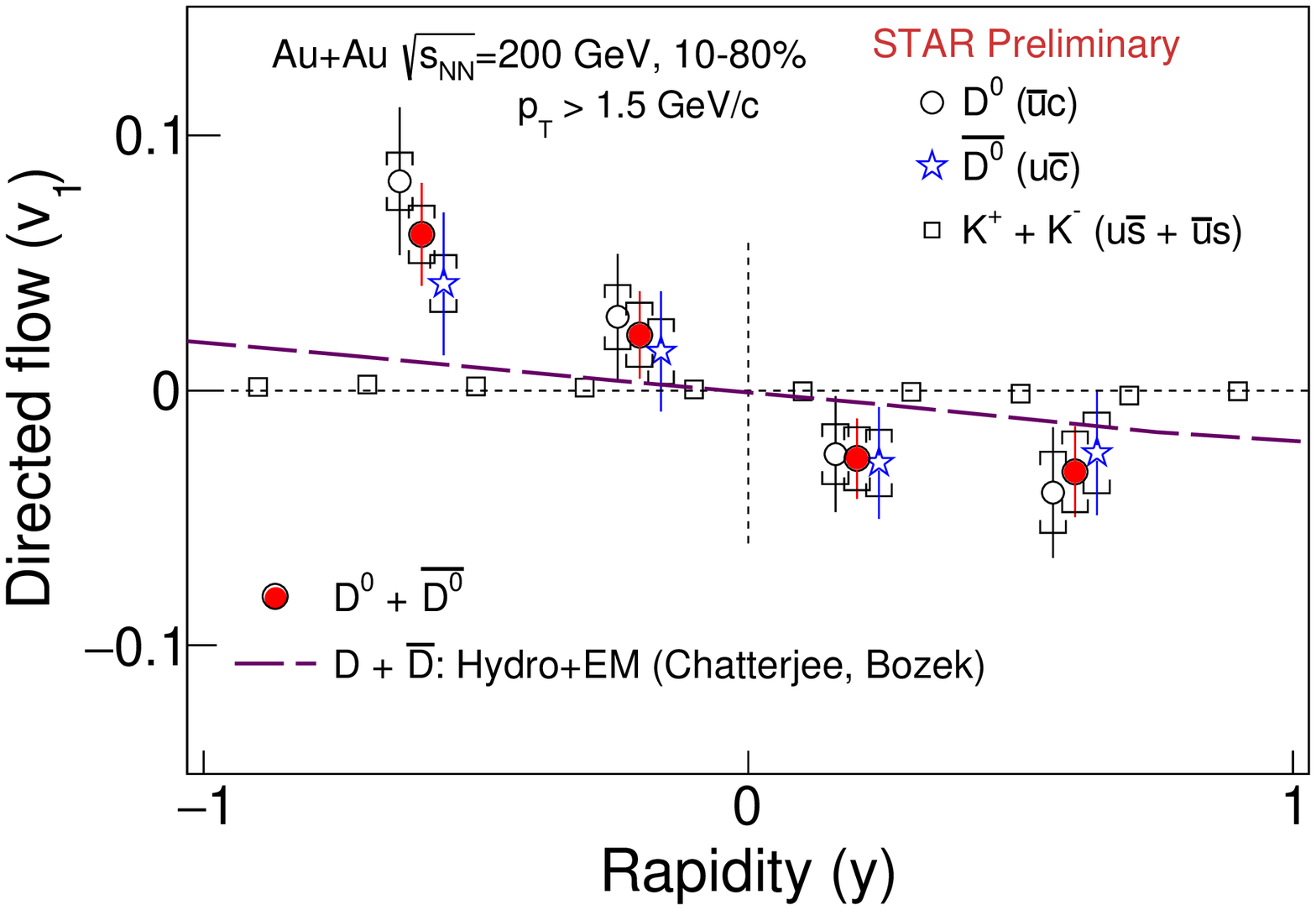}
\includegraphics[scale=0.35]{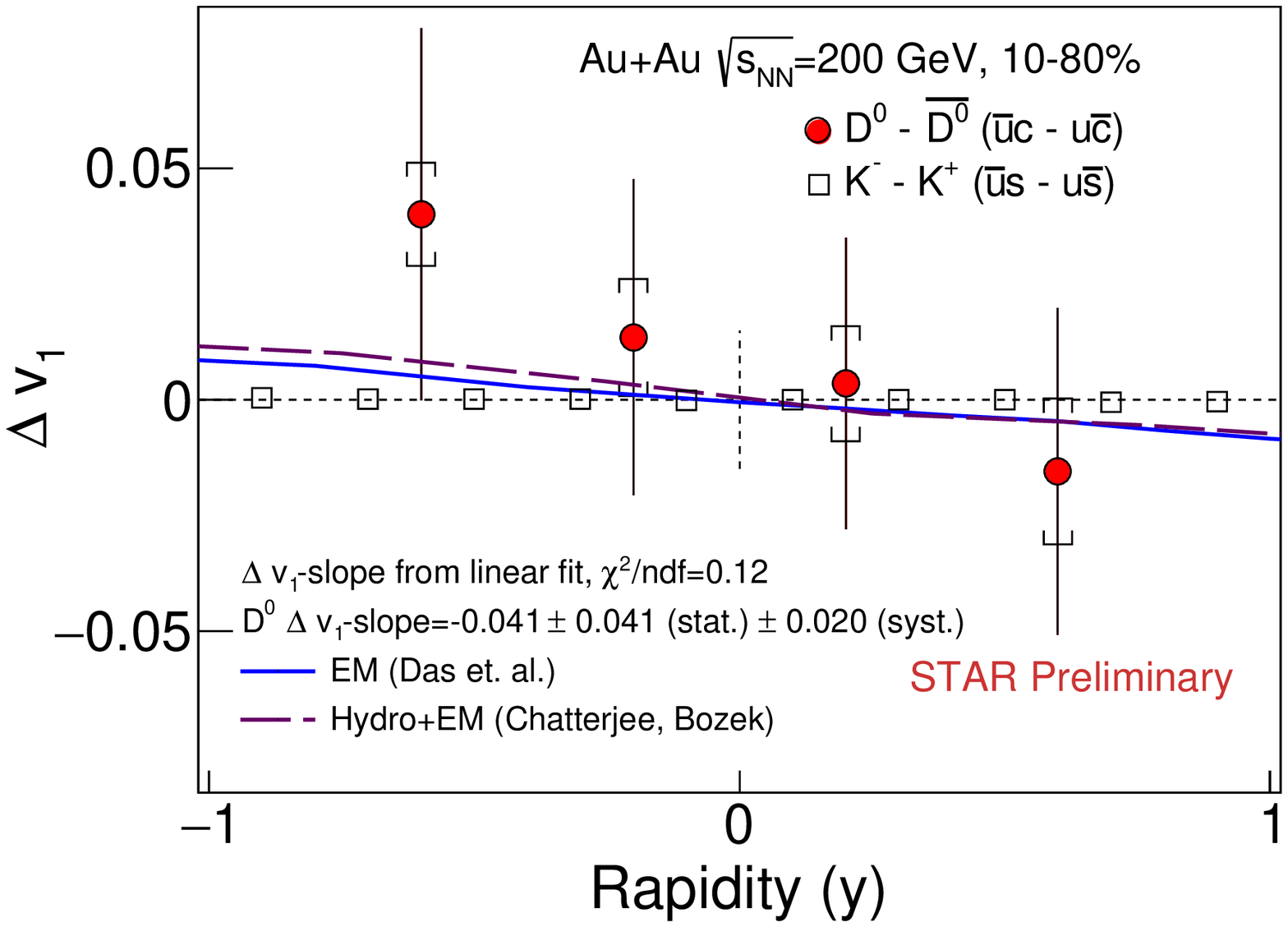}
\vspace*{-3mm}
\caption{(Color online) Left panel: Open circle and star markers
  present $D^{0}$ and $\overline{D^{0}}$ $v_{1}$ as a
  function of rapidity for $p_{T} > $1.5~GeV/$c$ in 10-80\% Au+Au
  collisions at $\sqrt{s_{\rm NN}}$ = 200~GeV, while the solid red circles
  are their average. The $D^{0}$ and $\overline{D^{0}}$ data points
  are shifted along the horizontal axis by $\pm$ 0.04 for visibility. Open
  black squares present $v_{1}(y)$ for charged kaons~\cite{STAR-BESv1}. The magenta dashed line 
  shows $v_{1} (y)$ from a hydro model calculation incorporating the initial electromagnetic
  field~\cite{chatt_bozek_dv1_2nd}. Right panel: Solid red symbols present the difference
between $D^{0}$ and $\overline{D^{0}}$, while the open black squares
are the difference between $K^{-}$ and $K^{+}$. The solid blue and
magenta dashed lines are the $\Delta~v_{1}$ prediction from Refs.~\cite{das_greco_dv1_em}
and ~\cite{chatt_bozek_dv1_2nd}, respectively. In both panels, the vertical bars and caps denote 
statistical and systematic uncertainties, respectively.
}
\label{fig1}
\end{center}
\end{figure}

The left panel in Fig.~\ref{fig1} presents the rapidity-odd directed flow for
$D^{0}$ and $\overline{D^0}$ mesons and their average
in 10-80\% central Au+Au collisions at $\sqrt{s_{\rm NN}}$ = 200~GeV with
$p_{T} >$ 1.5 GeV/$c$ using 2014 and 2016 data combined. The $v_1(y)$
slope for $D^{0}$ mesons is extracted by fitting the data with a
linear function constrained to pass through the origin. The choice of
using a linear function is driven by the limited $D^0$ statistics. The observed
$dv_{1}/dy$ for  $D^{0}$ and $\overline{D^0}$ is -0.102 $\pm$ 0.030 (stat.) $\pm$ 0.021 (syst.) and
 -0.061 $\pm$ 0.030 (stat.) $\pm$ 0.023 (syst.), respectively, while $dv_{1}/dy$ for their
 average is -0.081 $\pm$ 0.021 (stat.) $\pm$ 0.017 (syst.),
 corresponding to a 3$\sigma$ significance. The heavy flavor results are
 compared to the average of $K^+$ and $K^-$~\cite{STAR-BESv1}. The kaon $v_{1}$ 
 slope is obtained from a similar linear fit, and the fitted $dv_{1}/dy$ for kaons is 
 -0.0030 $\pm$ 0.0001 (stat.) $\pm$ 0.0002 (syst.). While the sign of
 $dv_1/dy$ is the same, the magnitude of $D^0$ $dv_{1}/dy$ is
 about 20 times larger (2.9$\sigma$ significance) compared to the kaon $dv_1/dy$. 
 A recent hydrodynamic model~\cite{chatt_bozek_dv1_prl} predicts that the drag
 from the initially tilted bulk can induce a relatively larger $v_{1}$ for heavy-flavor hadrons compared to light hadron species. Hence, the $D$ meson $v_{1}$ slope can be used to probe the initial thermal matter distribution in the longitudinal and transverse directions. Furthermore, the initial EM field can
 induce opposite $v_{1}$ for charm and anti-charm quarks. A hydrodynamic
 model calculation combined with initial EM fields
 suggests that the $v_{1}$ splitting due to the EM field can be smaller
 than the $v_{1}$ induced by the drag of the tilted bulk~\cite{chatt_bozek_dv1_2nd}. The
 dashed magenta line in the left panel of Fig.~\ref{fig1} represents the $v_{1}(y)$ prediction from this 
 hydrodynamic model incorporating the initial EM field~\cite{chatt_bozek_dv1_2nd}. The model 
 predicts the correct sign for both $D^{0}$ and $\overline{D^0}$ mesons, but the
 magnitude of $v_{1}$ is underestimated when using the particular choice of model
 parameters in Ref.~\cite{chatt_bozek_dv1_2nd}. The difference in
 $v_{1}$ ($\Delta$$v_{1}$) between the $D^{0}$ and 
 $\overline{D^0}$, predicted to be sensitive to the initial EM
 field, is presented in the right panel of Fig.~\ref{fig1}. The $D^{0}$
 results are compared with two model predictions, shown by solid
 blue~\cite{das_greco_dv1_em} and magenta dashed
 ~\cite{chatt_bozek_dv1_2nd} lines. The current precision of the data does not permit firm conclusions 
concerning the difference and ordering between the $D^{0}$ and
$\overline{D^0}$ mesons.
\vspace*{-3mm}
%%%%%%%%%%%%%% Fig. 1 %%%%%%%%%%%%%%%%%%%%%
\vspace*{-8mm}
%%%%%%%%%%%%%% Fig. 2 %%%%%%%%%%%%%%%%%%%%%%
\begin{figure}[!htb]
\begin{center}
\includegraphics[scale=0.38]{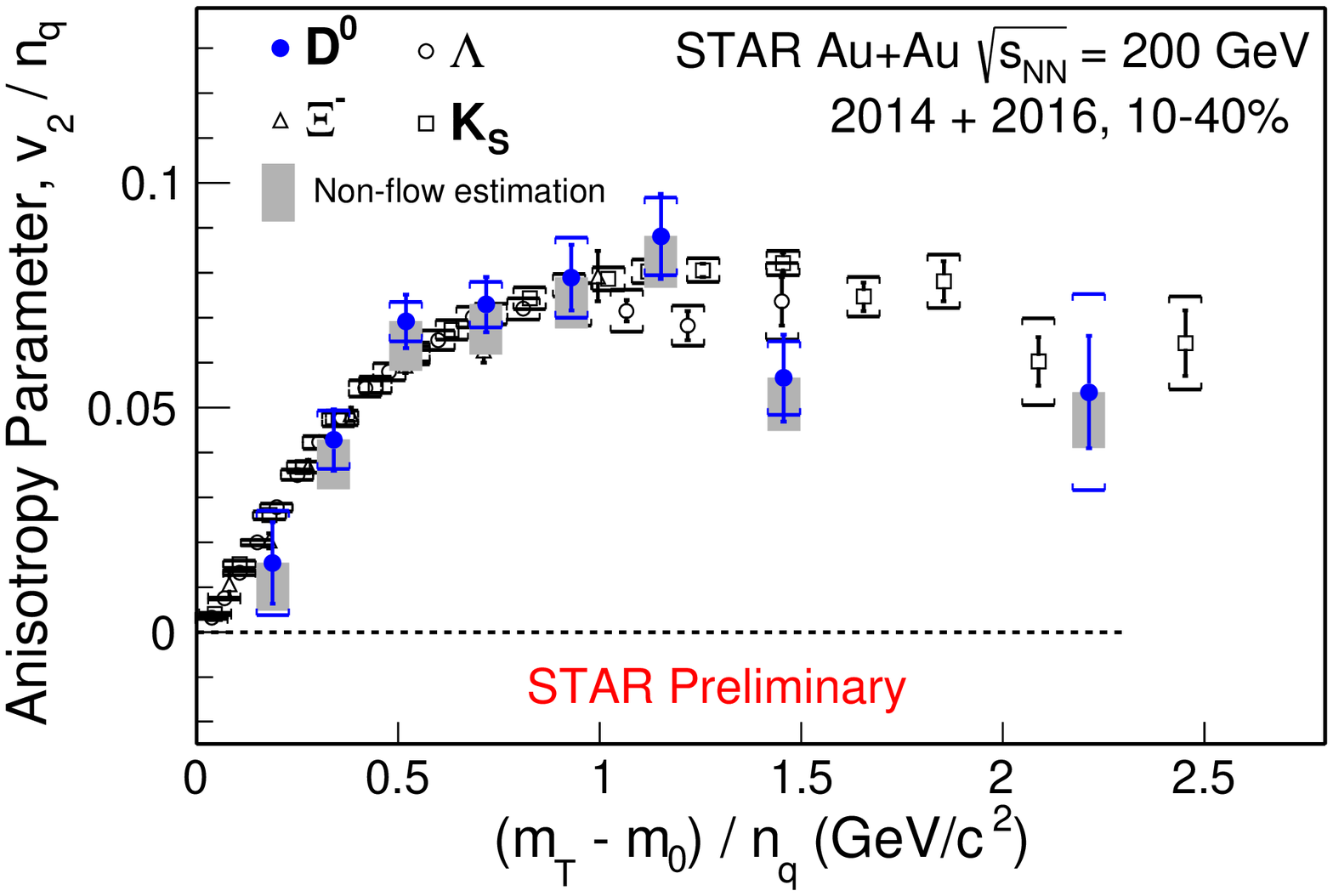}
\includegraphics[scale=0.35]{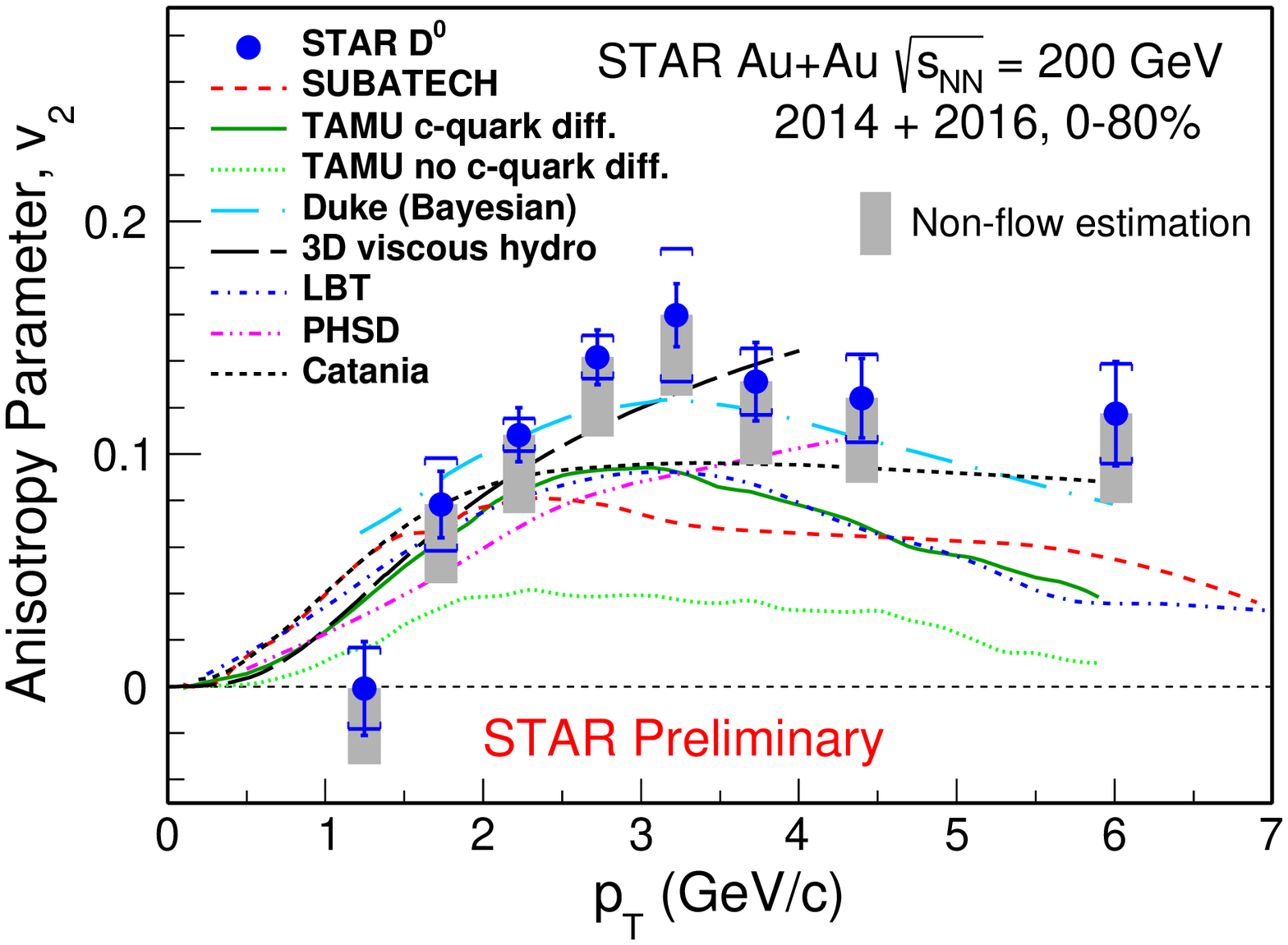}
\vspace*{-3mm}
\caption{(Color online) Left panel: $v_{2}/n_{q}$ as a function of
  $(m_{T} - m_{0})/n_{q}$ for $D^{0}$ and $\overline{D^{0}}$
  mesons combined in 10-40\% central Au+Au collisions at
  $\sqrt{s_{\rm NN}}$ = 200~GeV along with $K_{S}^{0}$, $\Lambda$, and
  $\Xi$~\cite{STAR-ks0-lam-v2}. Right panel: $v_{2}$ as a function of
  $p_{T}$ for $D^{0}$ and $\overline{D^{0}}$
  mesons combined in 0-80\% Au+Au collisions compared with model
  calculations~\cite{subatech, tamu, duke, 3d-hydro, lbt, phsd, catania}.} 
\label{fig2}
\end{center}
\end{figure}
%%%%%%%%%%%%%%%%%%%%%%%%%%%%%%%%%%%%%%%
\vspace*{-5mm}
The averaged $v_{2}(p_T)$ of $D^{0}$ and $\overline{D^{0}}$ mesons is measured in 0-10$\%$,
10-40$\%$ and 0-80$\%$ central Au+Au collisions at $\sqrt{s_{\rm NN}}$
= 200~GeV based on combined datasets recorded during 2014 and
2016. This provides about a 30$\%$ improvement in the statistical
precision compared to previously published results using 2014 data
alone~\cite{star_d0_v2_ncq}. The new results allow us to perform
improved NCQ-scaling tests. The blue solid markers in the left panel
of  Fig.~\ref{fig2} present the NCQ-scaled $v_{2}$ as a function of
NCQ-scaled transverse kinetic energy $(m_{T}-m_{0})$ for $D^{0}$
mesons in 10-40$\%$ central Au+Au collisions at $\sqrt{s_{\rm NN}}$ =
200~GeV. The results are compared to light hadron species, namely the
$K^{0}_{S}$ meson and the $\Lambda$ and $\Xi$
baryons~\cite{STAR-ks0-lam-v2}. The NCQ-scaled $D^{0}$ $v_{2}$ is
compatible within uncertainties with those of light hadrons for
$(m_{T}-m_{0})/n_{q} < $ 2.5 GeV/$c^2$. This observation suggests that
the charm quarks exhibit the same strong collective behavior as light-flavor quarks, and may be close to thermal 
equilibrium with the medium in Au+Au collisions at $\sqrt{s_{\rm NN}}$
= 200 GeV. The right panel in Fig.~\ref{fig2} presents the $D^{0}$ $v_{2}$ results in
0-80\% central Au+Au collisions, and compared to SUBATECH~\cite{subatech}, TAMU~\cite{tamu}, Duke~\cite{duke}, 3D viscous hydro~\cite{3d-hydro}, LBT~\cite{lbt}, PHSD~\cite{phsd},
and Catania~\cite{catania} model calculations. These models include
different treatments of the charm quarks interactions with the medium and they also
differ in their initial state conditions, QGP evolution,
hadronization, etc.  We have performed a statistical significance test
of the consistency between the data and each model, quantified by
$\chi^{2}$/NDF and the $p$ value. We have found that the TAMU model
without charm quark diffusion cannot describe the data, while the same
model with charm quark diffusion turned-on shows better agreement. All
the other models can describe the data in the measured $p_{T}$ region.
\section{Conclusion}
\label{conclusion}
In summary, we report on the first evidence for a non-zero rapidity-odd directed flow
of $D^{0}$ and $\overline{D^0}$ mesons in Au+Au 
collisions at $\sqrt{s_{\rm NN}}$ = 200~GeV in the 10-80$\%$
centrality class. The $dv_{1}/dy$ of the average of $D^{0}$ and
$\overline{D^0}$ mesons is -0.081 $\pm$ 0.021 $\pm$ 0.017, which is
significantly larger than that of the charged kaons having $dv_{1}/dy$
of -0.0030 $\pm$ 0.0001 $\pm$ 0.0002. Models indicate that the large $dv_1/dy$ of $D^0$ is
sensitive to the initially tilted source. However, the current
precision of the data is not sufficient to clearly determine the difference and ordering between $D^{0}$ and 
$\overline{D^0}$ mesons, which, according to models, is sensitive to the initial 
electromagnetic field. We also report on the elliptic flow as a function of $p_T$ for combined 
$D^{0}$ and $\overline{D^0}$ mesons combining 2014 and 2016 data
samples. The $D^{0}$ $v_{2}$ result suggests that the charm quark may be close to thermal equilibrium with
the medium. Furthermore, studies are now in progress in determining the $D^{0}$ 
$v_{2}$ in the peripheral collisions (40-80\%), with an enlarged pseudorapidity 
gap to reduce non-flow effects.
\bibliographystyle{elsarticle-num}
\bibliography{<your-bib-database>}

%%\end{linenumbers}
\end{document}